\documentclass[twocolumn,showpacs]{revtex4}
\usepackage{graphicx}
\usepackage{dcolumn}
\usepackage{bm}
\usepackage{amsmath}
\usepackage{amsfonts}
\usepackage{amssymb}
\usepackage{color}
\usepackage{epstopdf}
\providecommand{\U}[1]{\protect\rule{.1in}{.1in}}
\newcommand{\ket}[1]{|#1\rangle}
\newcommand{\bra}[1]{\langle#1|}

\newcommand{\modu}[1]{\vert\vec{#1}\vert}

\newcommand{\curvo}[1]{\left({#1}\right)}
\newcommand{\abso}[1]{\vert{#1}\vert}

\begin{document}
\title{Initial-state-dependent thermalization in open qubits}
\author{Andr\'es Vallejo, Alejandro Romanelli and Ra\'ul Donangelo}
\affiliation{Instituto de F\'{\i}sica, Facultad de Ingenier\'{\i}a\\
Universidad de la Rep\'ublica\\
C.C. 30, C.P. 11300, Montevideo, Uruguay}
\date{\today }

\begin{abstract}
We study, from a thermodynamic perspective, the equilibrium states
of a qubit interacting with an arbitrary environment of dimension $N\gg 2$.
We show that even in presence of memory about the initial state,
in some cases the qubit can be considered in a thermal state characterized
by an \textit{entanglement Hamiltonian}, which encodes the effects of the
environment, and an initial-state-dependent \textit{entanglement temperature}
that measures the degree of entanglement generated between the system and its
environment.
Geometrical aspects of the thermal states are studied, and the results are
confirmed for the concrete case of the Quantum Walk on the Line.
\end{abstract}

\maketitle

\section{Introduction}
\textit{Quantum thermodynamics} tries to account for the emergence, from
the principles of Quantum Mechanics, of the thermodynamic behavior observed
in macroscopic systems.
This is not an easy task, since the unitary evolution of isolated quantum
systems implies that the expected values of the corresponding physical
quantities oscillate in time, and, as a consequence, the system is in
general out of equilibrium \cite{Schulman}.
For a general review of this problem see Ref.\cite{Gemmer}, and more recent
discussions can be found in Refs.\cite{Goold,Parrondo}.

Recent theoretical work has considered the Eigenstate Thermalization
Hypothesis, showing that quantum systems can indeed thermalize
\cite{Polkovnicov}.
This hypothesis provides a framework connecting microscopic models and
macroscopic phenomena, based on the notion of highly entangled quantum states.
The role of quantum entanglement in the emergence of statistical mechanics
in the quantum regime has been experimentally studied in Ref.\cite{Kaufman}.
This work shows that the entanglement entropy can be measured, playing a
similar role as that of the thermal entropy in the classical thermalization
processes.
This validates the use of Statistical Physics for the measurement of
local observables.

When we consider open quantum systems, we observe that in some cases the
interaction with the environment may lead the system to an equilibrium state,
despite the unitarity of the global dynamics.
Although in general it is not an equilibrium \textit{stricto sensu}, but rather
a situation in which the system spends most of the time in the neighborhood of
an average state, an attempt to conciliate thermodynamics with such kind of
systems is therefore possible \cite{Romanelli1, Romanelli2, Novotny, schliemann}.
In particular, it has been proved that a large energy-level occupation of the
environment in the initial state establishes strong bounds to the distance
between the reduced state of the subsystem and its corresponding time-average,
which assures equilibration if the mentioned condition is satisfied \cite{Linden}.
Moreover, the canonical typicallity of reduced states of bipartite quantum systems
has been proved, showing the ubiquity of the thermal behavior, even at the
nanoscale \cite{Goldstein}, \cite{Popescu}.  \\

In this context, we explore the equilibrium states of an open qubit, which,
due to interaction with the environment, evolves to an initial-state-dependent
equilibrium state.
We find that in some cases, this dependence can be factorized in such a way
that the reduced density operator (RDO) adopts the form of thermal state for
some fixed operator that we call \textit{entanglement Hamiltonian}, which
depends on the relevant parameters of the global dynamics.
Meanwhile, the corresponding \textit{entanglement temperature} encloses the
dependence on the initial state and can be interpreted as a measure of the
entanglement generated during the process.\

This paper is organized as follows.
In section II, we introduce the concepts of entanglement Hamiltonian and
entanglement temperature, notions that will allow us to define the thermal
states in a proper way.
Then, we explore some of their properties and find a property that allows
to identify which of the asymptotically equilibrated initial-state-dependent
reduced operators correspond to thermal states. In section III, we study this
class of thermal states from a geometric perspective, finding their location
on the Bloch sphere.
In section IV we find the thermal states associated to the chirality
degrees of freedom for a Quantum Walk on the Line.
Finally, some remarks and perspectives are discussed in section V.\

\section{RDO and thermal states}
Let us consider an isolated  bipartite quantum system composed of a
qubit $S$ in thermal contact with an arbitrary environment $E$, so
that the state space for the composed system
is
\begin{equation}\label{eqn:jilbertspace}
\mathcal{H}=\mathcal{H}_{_{E}}\otimes \mathcal{H}_{_{S}}.
\end{equation}
 Let us suppose there is no initial correlation between the systems, so
 the initial state is a product state, that can be written as
\begin{equation}\label{eqn:initial}
\ket{\Psi(0)}=\sum_{n}C_{n}\ket{n}\otimes \ket{\phi_0},
\end{equation}
with
\begin{equation} \label{eqn:phi0}
\ket{\phi_0}={\cos\frac{\gamma}{2}\ket{+} +e^{i\varphi}\sin
\frac{\gamma}{2}\ket{-}},
\end{equation}
where
\begin{itemize}
\item $\lbrace \ket{+} , \ket{-}\rbrace $ and $\lbrace
{\ket{n}}\rbrace$ are some orthonormal bases in $\mathcal{H}_{_{S}}$ and
$\mathcal{H}_{_{E}}$ respectively,
\item $\gamma$ and $\varphi$ define the point over the Bloch sphere
associated to the initial state of the qubit,
\item $\lbrace C_{n}\rbrace$ are the amplitudes that define the initial
occupation of the bath and satisfy the normalization condition
$\sum_{n}\vert C_{n}\vert^{2}=1$.
\end{itemize}

If the system equilibrates \textit{on average} via interaction with
the environment, the equilibrium state is described by the
time-averaged RDO \cite{Nielsen}
\begin{equation}\label{eqn:timeaveraged}
\begin{split}
\overline{\rho_{_{S}}}(\lbrace C_{n}\rbrace,\gamma,\varphi)&=
\langle\rho_{_{S}}(t,\lbrace C_{n}\rbrace,\gamma,\varphi)\rangle_{t} \\
&=\lim_{\tau\to\infty}\frac{1}{\tau}\int_{0}^{\tau}
\rho_{_{S}}(t,\lbrace C_{n}\rbrace,\gamma,\varphi)dt,\\
\end{split}
\end{equation}
where $\rho_{_{S}}=tr_{_{E}}\rho_{_{SE}}=tr_{_{E}}
\ket{\Psi}\bra{\Psi}$. Without loss of generality,
$\overline{\rho_{_{S}}}$ can be written as
\begin{small}
\begin{equation}\label{eqn:ARDM}
\overline{\rho_{_{S}}}(\lbrace C_{n}\rbrace,\gamma,\varphi)=
\begin{pmatrix}
{\dfrac{1}{2}+a}&
             {b}\\
             {b^{*}}&
{\dfrac{1}{2}-a}
\end{pmatrix}.
\end{equation}
\end{small}
where $a$ and $b$ depend on the initial conditions
$\lbrace C_{n}\rbrace$, $\gamma$ and $\varphi$.
In a few interesting cases, the limit (\ref{eqn:timeaveraged}) coincides
with the asymptotic limit of the RDO as $t\to\infty$ \cite{Romanelli1}.
But in most of cases, the notion of equilibration on average (or in another
relevant sense) is necessary since, due to the Quantum Recurrence Theorem,
in general the asymptotic limit of local operators does not exist \cite{Bocchieri}.\\

Now we will investigate if, despite the initial state dependence,
there are situations in which the system can be considered in
thermal equilibrium, {\it i.e.}, if there exists an operator
$H_{ent}$, defined on $\mathcal{H_S}$, and a real number $\beta$
such that $\overline{\rho_{_{S}}}$ adopts the form
\begin{equation}\label{eqn:termal}
\overline{\rho_{_{S}}}=\dfrac{e^{-\beta H_{ent}}}{tr(e^{-\beta H_{ent}})}.
\end{equation}

Note that the operator $H_{ent}$ may not coincide with the free Hamiltonian,
due to environment effects.
The Hamiltonian $H_{ent}$ that governs the dynamics in the asymptotic
regime cannot (or, at least, should not) be initial-state-dependent.
So we will center our attention in investigating if it is possible to
factorize the exponent in Eq.(\ref{eqn:termal}) in such a way that the
initial state dependence appears as a scalar factor, that we will interpret
as the entanglement temperature, multiplied by a initial-state-independent
operator, which plays the role of an entanglement Hamiltonian.
When such a factorization is possible, the entanglement Hamiltonian will be
well-defined except for an additive multiple of the identity (related to an
irrelevant energy shift), and a multiplicative factor that represents an also
irrelevant change of scale in the temperature.
Then, we will have that systems in different initial states, governed by the
same entanglement Hamiltonian, will thermalize at different inverse entanglement
temperatures $\beta$.\\
As we will see,  the aforementioned factorization is only
possible for a small set of initial-state-dependent RDOs, that we
will call \textit{initial-state-dependent thermal states}.

The parameter $T_{ent}=1/\beta$ is, a priori,
not related to the thermodynamic temperature and will be denoted
\textit{entanglement temperature}.
It is easy to show the following properties of the thermal states defined
in Eq.(\ref{eqn:termal}):
\begin{itemize}
\item If ${\lambda_{j}}$ are the natural populations of the system S and
${E_{j}}$ its energy eigenvalues, then $\delta Q = T_{ent}dS_{vN}$, where
$\delta Q= \sum_{j}E_{j}d\lambda_{j}$ is the \textit{heat} exchanged with
the environment, and $S_{vN}=-\sum_{j}\lambda_{j}\log{\lambda_{j}}$ is
the von Neumann entropy \cite{Gemmer}.
This result, valid for infinitesimal transitions between thermal states
emphasizes the similarity between entanglement thermodynamics and classical
thermodynamics.
\item $T_{ent}=0$ for pure reduced states and $T_{ent}=\infty$ for the maximally
mixed state.
In fact, $T_{ent}$ is a growing function of $S_{vN}$, so the entanglement
temperature is a measure of the entanglement generated between the system
and its environment.\end{itemize}

We note that the positivity of the RDO
$\overline{\rho_{_{S}}}$, implies that it is always possible to express it
in the exponential form of Eq.(\ref{eqn:termal}), so, \textit{a priori},
every reduced state is, potentially, a thermal state.
In fact, there is an infinite number of ways of selecting $\beta$ and
$H_{ent}$ for a given state, so the system can be considered in equilibrium
at any temperature, depending on the choice of $H_{ent}$.

In order to obtain an explicit condition for such factorization, let
us consider the basis $\lbrace \ket{\psi^{+}}$, $\ket{ \psi^{-}}
\rbrace $ that diagonalizes $\overline{\rho_{_{S}}}$. The functional
relation between $\overline{\rho_{_{S}}}$ and $H_{ent}$ implies that
the latter will also have a diagonal expression in that basis. If we
denote by $\lbrace\varepsilon$, $-\varepsilon \rbrace$ the
eigenvalues of $H_{ent}$, and by
$\lbrace\lambda^{+}$,$\lambda^{-}\rbrace$ the corresponding natural
populations, we have
\begin{equation}
H_{ent}^{(diag)}=\begin{pmatrix}
{\varepsilon}&{0}\\
{0}&{-\varepsilon}
\end{pmatrix}, \;\;
\overline{\rho_{_{S}}}^{(diag)}=\begin{pmatrix}
{\lambda^{+}}&{0}\\
{0}&{\lambda^{-}}
\end{pmatrix}.
\end{equation}

Now, observe that if  ${\overline{\rho_{_{S}}}}^{(diag)}$ can be obtained
from $\overline{\rho_{_{S}}}$ through the change of basis
\begin{equation}
{\overline{\rho_{_{S}}}}^{(diag)}=Q^{\dag}\overline{\rho_{_{S}}}Q,
\end{equation}
\noindent then, via the inverse transformation, we can find the
explicit form of $H_{ent}$ in the base $\lbrace \ket{+} ,
\ket{-}\rbrace $:
\begin{equation} \label{eqn:Heff cb}
H_{ent}=QH_{ent}^{(diag)}Q^{\dag}.
\end{equation}
Let us implement the described procedure.
We first find the eigenvalues of (\ref{eqn:ARDM}):
\begin{small}
\begin{equation}\label{eqn:eigenvector1}
\lambda^\pm= \dfrac{1}{2}\pm\sqrt{a^{2}+\vert b\vert^{2}},
\end{equation}
\end{small}
and the corresponding eigenvectors:
\begin{small}
\begin{equation}\label{eqn:psi}
\ket{\psi^\pm}= \begin{pmatrix}
{\psi_1^\pm}\\
\vspace{0.2cm} {\psi_2^\pm}
\end{pmatrix},
\end{equation}
\end{small}
where
\begin{small}
\begin{equation}\label{eqn:psi2}
\begin{cases}
\psi_1^\pm=\dfrac{\abso{b} \left(a\pm\sqrt {a^{2}+\abso{b}^{2}}\right)}{b^{*}\sqrt{\left( a\pm\sqrt{a^{2}+
\abso{b}^{2}}\right)^{2}+\abso{b}^{2}}},\\
\vspace{0.2cm} \psi_2^\pm=\dfrac{\abso{b}}{\sqrt{\curvo{ a\pm\sqrt{a^{2}+\abso{b}^{2}}}^{2}+\abso{b}^{2}}}.
\end{cases}
\end{equation}
\end{small}
So the change of basis matrix that simultaneously diagonalizes the
operators $\overline{\rho_{_{S}}}$ and $H_{ent}$ is
\begin{small}
\begin{equation}
Q=\begin{pmatrix} {\psi_1^+} &{\psi_1^-}
\vspace{0.2cm}\\
{\psi_2^+}&{\psi_2^-}
\end{pmatrix}.\
\end{equation}
\end{small}
Now, employing Eq.(\ref{eqn:Heff cb}) we obtain the following
expression for the entanglement Hamiltonian in the base $\lbrace
\ket{+}, \ket{-}\rbrace $:
\begin{small}
\begin{equation}\label{eqn:effective hamiltonian}
H_{ent}=QH^{(diag)}_{ent}Q^{\dag} =\varepsilon\begin{pmatrix}
{\dfrac{a}{\sqrt{ a^{2}+\abso{b}^{2}}}}&
{\dfrac{b}{\sqrt{ a^{2}+\abso{b}^{2}}}}\\
{\dfrac{b^{*}}{\sqrt{ a^{2}+\abso{b}^{2}}}}& {\dfrac{-a}{\sqrt{
a^{2}+\abso{b}^{2}}}}
\end{pmatrix}.
\end{equation}
\end{small}
Equation (\ref{eqn:effective hamiltonian}) shows an evident dependence on the initial state through the
coefficients $a$ and $b$. We then conclude that, in general, it is not possible to express the equilibrium
state in the form of a thermal state (Eq.(\ref{eqn:termal})) with an universal, {\it i.e.} valid for any
initial condition,  entanglement Hamiltonian. However, there is a particular relation between the parameters
$a$ and $b$ that makes $H_{ent}$ independent of the initial state. Note that, if a constant $\kappa$
independent of $\gamma$, $\varphi$ and $\lbrace C_{n}\rbrace$ can be found such that $b=\kappa a$, and taking
into account that $a\in R$, we have
\begin{equation}\label{eqn:Hent}
 H_{ent} =\dfrac{\varepsilon}{\sqrt{ 1+\abso{\kappa}^{2}}}
 \begin{pmatrix}{1}&
{\kappa}\\
{\kappa^{*}}& {-1}
\end{pmatrix}.
\end{equation}

Reciprocally, if Eq.(\ref{eqn:effective hamiltonian}) is independent
of the initial state, we can write
 \begin{small}
 \begin{equation}
 \varepsilon\begin{pmatrix}
{\dfrac{a}{\sqrt{ a^{2}+\abso{b}^{2}}}}&
{\dfrac{b}{\sqrt{ a^{2}+\abso{b}^{2}}}}\\
{\dfrac{b^{*}}{\sqrt{ a^{2}+\abso{b}^{2}}}}& {\dfrac{-a}{\sqrt{
a^{2}+\abso{b}^{2}}}}
\end{pmatrix}=\begin{pmatrix}
{\eta_{1}}&
{\eta_{2}}\\
{\eta^{*}_{2}}& {\eta_{1}}
\end{pmatrix},
 \end{equation}
 \end{small}
where $\eta_{1}$ and $\eta_{2}$ do not depend on
$\lbrace C_{n}\rbrace$, $\gamma$, $\varphi$.
Then, we have
 \begin{equation}
 b/a= \dfrac{\dfrac{b}{\sqrt{ a^{2}+\abso{b}^{2}}}}{\dfrac{a}{\sqrt{ a^{2}+\abso{b}^{2}}}}=\dfrac{\eta_{2}}{\eta_{1}}=\kappa ,
 \label{bovera}
 \end{equation}
where $\kappa$ is a constant.

The previous analysis can be summarized in the following proposition:
\newtheorem{prop}{Proposition}
\begin{prop}\label{propos1}
A time-averaged RDO of the type of Eq.\text{(\ref{eqn:ARDM})} describing
an equilibrated qubit is an initial-state-dependent thermal state if, and
only if there exists a set of initial conditions such that:
\begin{equation}\label{eqn:condition}
b(\lbrace C_n\rbrace,\gamma ,\varphi)=\kappa a(\lbrace C_n\rbrace,\gamma ,\varphi),
\end{equation}
\noindent where  $\kappa \in C$ is a constant independent of the initial state.
In that case, the entanglement Hamiltonian in the basis $\lbrace
\ket{+}, \ket{-}\rbrace$ takes the form
\begin{equation}
H_{ent} =\dfrac{\varepsilon}{\sqrt{ 1+\vert \kappa \vert^{2}}}
\begin{pmatrix}
{1}&{\kappa}\\
{\kappa^{*}}&{-1}
\end{pmatrix},
\end{equation}
except for an arbitrary multiplicative factor, or an additive multiple of
the identity, which do not lead to relevant physical consequences.
\end{prop}

\section{The role of the entanglement Hamiltonian: geometry of thermal states}
In general, we expect that the dimensionless quantity $\kappa$ that defines
the entanglement Hamiltonian can be constructed from the parameters involved
in the global Hamiltonian that characterize the system and its interaction with
the environment, such as coupling constants or characteristic frequencies. \\
In particular, a simple consequence of the restriction
$b(C_n,\gamma,\varphi)=\kappa a(C_n,\gamma ,\varphi)$ arises by analyzing the
location of the thermal states on the Bloch sphere.

Because of Proposition (\ref{propos1}), we know that the RDO of an
initial-state-dependent thermal state can be written in the form
\begin{small}
\begin{equation}\label{eqn:prop1}
\overline{\rho_{_{S}}}= \begin{pmatrix} {\dfrac{1}{2}+a}&{\kappa a}\\
{\kappa^{*}a}& {\dfrac{1}{2}-a}
\end{pmatrix},
\end{equation}
\end{small}
\noindent where the dependence on the initial state is totally
included in $a$. On the other hand, in terms of the Bloch vector
components, $\vec{B}=(B_{1}, B_{2}, B_{3})$, representing the state,
the expression for this operator is \cite{Nielsen}
\begin{small}
\begin{equation}
\overline{\rho_{_{S}}}= \dfrac{1}{2}\begin{pmatrix}
{1+B_{3}}&{B_{1}-iB_{2}}\\
{B_{1}+iB_{2}}& {1-B_{3}}
\end{pmatrix}.
\end{equation}
\end{small}
Comparing both expressions we have that
\begin{equation}\label{eqn:vector de bloch}
\vec{B}=2a \vec{v},
\end{equation}
with
\begin{equation}
\vec{v}=\left({\rm {Re}}(\kappa),-{\rm {Im}}(\kappa),1\right),
\end{equation}
where ${\rm {Re}}(\kappa)$ and ${\rm {Im}}(\kappa)$ denote the real and imaginary parts of $\kappa$. Note
that the direction of the Bloch vector is completely defined by the quantity $\kappa$, while the initial
state only plays a role in fixing its modulus. Given that $\kappa$ only depends on the relevant parameters
involved in the global dynamics, which are fixed in advance, we conclude that the entanglement Hamiltonian
fixes, through the parameter $\kappa$, the diameter of the Bloch sphere that corresponds to the locus of the
thermal states. In fact, observe that the entanglement Hamiltonian (\ref{eqn:Hent}) can be expressed as
\begin{equation}
H_{ent}=\frac{\varepsilon}{\sqrt{1+\vert\kappa\vert^{2}}} \left(\vec{v}.\vec{\sigma}\right),
\end{equation}
where $\vec{\sigma}=\left(\sigma_x,\sigma_{y},\sigma_{z}\right)$ is
the vector whose components are the Pauli matrices.
So the vector $\vec{v}$ can be interpreted as an effective magnetic
field that selects the privileged direction along which the spin will
relax due to interaction with the environment.

\begin{figure}[]
\centering
\includegraphics[trim= 300 80 -50 0, scale=0.5, clip]{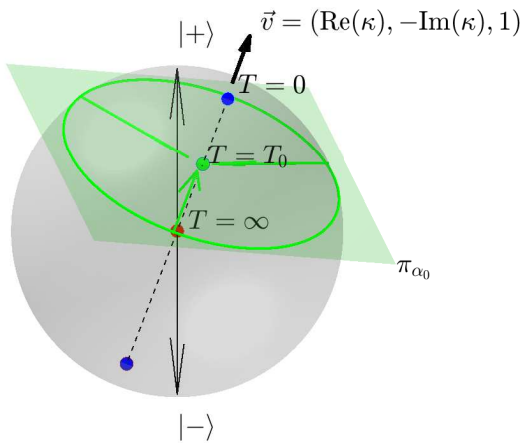}
\caption{(Color online) Level plots of the entanglement temperature over the
    Bloch sphere for a time-averaged RDO whose Kraus operators are orthogonal
    projectors.
The temperature $T_0$ (green point) corresponds to the asymptotic limit of all initial pure states located on
the intersection of the plane ${\rm {Re}}(\kappa)x-{\rm
{Im}}(\kappa)y+z=\sqrt{1+\modu{\kappa}^{2}}\cos\alpha_{0}$ with the sphere, as shown by the arrows. Also
represented are the minimum temperature states $T=0$ (blue points), and the $T=\infty$ state (red).}
\label{fig1}
\end{figure}

As an example, let us analyze the case in which $\rho_{_{S}}$
possesses an asymptotic limit $\rho_{_{S}}^{\infty}$ that admits a
Kraus  representation in terms of orthogonal projectors, {\it i.e.}
\begin{small}
\begin{equation}
\rho_{_{S}}^{\infty}=\sum_{j} M_j\rho_{_{S}}(0)M_j^\dag,
\end{equation}
\end{small}
where $M_{j}=\ket{\psi_{j}}\bra{\psi_{j}}$ and
$\bra{\psi_{i}}\psi_{j}\rangle=\delta_{ij}$, $j=1,2$.
This could seem rather artificial, but in the following section we
provide an example of a relevant quantum system verifying  such
conditions.\\
It is clear that in this case, the time-averaged RDO,
$\overline{\rho_{_{S}}}$, coincides with the asymptotic limit,
$\rho_{_{S}}^{\infty}$. If $\rho_{_{S}}(0)=
\ket{\phi_0}\bra{\phi_0}$ then
\begin{equation} \label{eqn:eigenvalues2}
\overline{\rho{_{S}}}\ket{\psi_{1,2}}= \vert \bra{\phi_0}
\psi_{1,2}\rangle \vert^{2}\ket{\psi_{1,2}},
\end{equation}
so the kets $\ket{\psi_{1,2}}$ are eigenvectors of
$\overline{\rho_{_{S}}}$, with corresponding eigenvalues
$\abso{\langle \phi_0\ket{\psi_{1,2}}}^{2}$, so it is clear that
\begin{equation}\label{psimas}
\ket{\psi_{1,2}}=\ket{\psi^{\pm}},
\end{equation}
where $\ket{\psi^{\pm}}$ are given by (\ref{eqn:psi}). Using Eqs.
(\ref{eqn:eigenvalues2}) and (\ref{psimas}) we have that
\begin{equation} \label{eqn:eigenvalues3}
\lambda^{+}-\lambda^{-}= \vert
\bra{\phi_0}\psi^{+}\rangle\vert^{2}-\vert
\bra{\phi_0}\psi^{-}\rangle\vert^{2},
\end{equation}
and using Eqs.(\ref{eqn:eigenvector1}), (\ref{eqn:psi2}),
(\ref{eqn:condition}) and (\ref{eqn:eigenvalues3}), it is
straightforward to show that
\begin{small}
\begin{equation}\label{eqn:a}
a=\dfrac{\cos\alpha}{2\sqrt{1+\abso{\kappa}^{2}}},
\end{equation}
\end{small}
where
\begin{equation} \label{cosenoalpha}
 \cos\alpha\equiv
 \frac{{\rm {Re}}(\kappa) \sin\gamma \cos \varphi-
 {\rm {Im}}(\kappa)\sin\gamma\sin\varphi+\cos\gamma}{\sqrt{1+\abso{\kappa}^{2}}}.
\end{equation}
It is interesting to point out that Eq.(\ref{cosenoalpha}) can
be expressed as
\begin{equation} \label{cosenoalpha2}
 \cos\alpha=\vec{B_{0}}.\frac{\vec{v}}{|\vec{v}|},
\end{equation}
where
$\vec{B}_{0}=(\sin\gamma\cos\varphi,\sin\gamma\sin\varphi,\cos\gamma)$
is the initial Bloch vector.
From Eq.(\ref{cosenoalpha2}) we see that $\alpha$ is the angle between
$\vec{B}_{0}$ and $\vec{v}$.
Finally, the expression for $\overline{\rho_{_{S}}}$ is
\begin{small}
\begin{equation}
\overline{\rho_{_{S}}}= \frac{1}{2}\begin{pmatrix}
{1+\dfrac{\cos\alpha}{\sqrt{1+\abso{\kappa}^{2}}}}&
{{\dfrac{\kappa\cos{\alpha}}{\sqrt{1+\abso{\kappa}^{2}}}}}\\
{{\dfrac{\kappa^{*}\cos\alpha}{\sqrt{1+\abso{\kappa}^{2}}}}}&
{1-\dfrac{\cos\alpha}{\sqrt{1+\abso{\kappa}^{2}}}}
\end{pmatrix} .
\end{equation}
\end{small}
The entanglement temperature can be obtained from the eigenvalues of
$\overline{\rho_{_{S}}}$, so we can construct the isotemperature
contour lines on the Bloch sphere (see Fig.\ref{fig1})
considering
\begin{small}
\begin{equation}\label{Tlambdas}
T_{ent}=\dfrac{2\varepsilon} {\log\left(\dfrac{\lambda^{+}}
{\lambda^{-}}\right)}=\dfrac{-\varepsilon}{\log\left[\tan\alpha/2\right]},
\end{equation}
\end{small}
which implies that these lines correspond to constant values of the angle
$\alpha$.
Expressing this condition in Cartesian coordinates,
$x=\sin\gamma\cos\varphi$, $y=\sin\gamma\sin\varphi$,
$z=\cos\gamma$, we obtain
\begin{small}
\begin{equation}\label{eqn:planes}
{\rm {Re}}(\kappa)x-{\rm {Im}}(\kappa)y+z= \sqrt{1+\abso{\kappa}^{2}}\cos{\alpha}.
\end{equation}
\end{small}
Eq.(\ref{eqn:planes}) represents a family of parallel planes, which are orthogonal to the vector $\vec{v}$
(see Fig.\ref{fig1}). The intersection of these planes with the Bloch surface produces a family of concentric
circumferences, each of which characterized by a specific value of asymptotic entanglement temperature. For
the particular initial states on the equatorial plane ${\rm {Re}}(\kappa)x-{\rm {Im}}(\kappa)y+z=0$
($\alpha=\pi/2$), the matrix expression for the RDO adopts the maximally mixed form
\begin{equation}
\overline{\rho_{_{S}}}=\frac{1}{2}\begin{pmatrix} {1}&{0}\\
{0}&{1}
\end{pmatrix}.
\end{equation}
As we move over on the northern hemisphere's surface towards the pole
defined by the vector $\vec{v}$ ($\alpha=0$), the temperature decreases
from $T_{ent}\rightarrow +\infty$ near the equatorial plane,
reaching its minimum value $T_{ent}=0$ at the North pole in Fig.\ref{fig1}.
Similarly, as we move towards the South pole, defined by the vector $\vec{v}$
($\alpha=\pi$), the entanglement temperature increases from $T_{ent}=-\infty$
at the equator to $T_{ent}=0$ at the pole.

In what refers to the location of a particular thermal state on the
Bloch sphere, it is clear that its Bloch vector is orthogonal to the
level plots of the entanglement temperature. The distance
between the center of the sphere $\vec{O}=(0,0,0)$ and the plane
$\pi_{\alpha_0}$ of Eq.(\ref{eqn:planes}) is given by
\begin{equation}
d(\vec{O},\pi_{\alpha_{0}})=
\frac{\abso{\sqrt{1+\abso{\kappa}^{2}}\cos\alpha_{0}}}{\sqrt{1+\kappa^{2}}}=\abso{\cos\alpha_{0}}.
\end{equation}
\noindent While the norm of $\vec{B}$, Eq.(\ref{eqn:vector de
bloch}) is
\begin{equation}
\vert \vec{B} \vert=2\vert a \vert \sqrt{1+\kappa^{2}},
\end{equation}
where $a$ is given by Eq.(\ref{eqn:a}). Thus, we have
\begin{equation}
\begin{split}
\modu{B}=\abso{\cos\alpha_{0}}= d(\vec{O} ,\pi_{\alpha_{0}}).
\end{split}
\end{equation}
So, we conclude that the thermal state is located at the center of the
corresponding entanglement temperature level plot, see Fig.\ref{fig1}.

\section{Example: Thermalization in the Quantum Walk on the line}

The Quantum Walk (QW) is the quantum analog of the classical random walk, and
it has been studied from multiple perspectives given the wealth of its applications
\cite{travaglione}, \cite{Renato}, \cite{Grover}.
In particular, the relation between the asymptotic coin-position entanglement
and the initial conditions of the QW has been investigated by several authors
\cite{Nayak,Carneiro,Abal,Annabestani,Omar,Pathak,
Petulante,Venegas,Endrejat,Ellinas1,Ellinas2,Maloyer}.
In this context, we shall emphasize the thermodynamical aspects, as an
application of the previous section.
We begin by noting that the mathematical structure behind the QW allows us
to consider the chirality degrees of freedom as a qubit $S$, described by a
two-dimensional Hilbert space $\mathcal{H}_{S}$, in interaction with a
thermal bath, represented by the walker's Hilbert space $\mathcal{H}_{n}$.
Then, we will show that for an adequate initial state of the bath, the
asymptotic reduced state of the qubit is an initial-state-dependent thermal
state, which will allow to illustrate the results obtained in the
previous section and,  in this context, give a physical interpretation to
the parameter $\kappa$.

First, we briefly review the QW dynamical equations in discrete
time. The system evolves under successive applications of the
evolution operator $U$, acting on
$\mathcal{H}_{n}\otimes\mathcal{H}_{S}$
\begin{equation}
\label{U}
U= \mathcal{T}(I_{n}\otimes U_{\theta}),
\end{equation}
where $U_{\theta}$ is an unitary evolution operator in 2D,
describing the quantum coin and parameterized by the \textit{coin bias}
parameter $\theta$,
\begin{equation}
U_{\theta}=\begin{pmatrix}
{\cos\theta}&{\sin\theta}\\
{\sin\theta}&{-\cos\theta}
\end{pmatrix}.
\end{equation}
\noindent This particular parameterization of $U_{\theta}$ is sufficient to
display all the possible evolutions of the QW.

In Eq.(\ref{U}) $\mathcal{T}$ is the conditional translation operator
\begin{equation}
\mathcal{T}=\sum_{n}\vert n+1\rangle \langle n\vert\otimes \vert +\rangle
\langle +\vert+\sum_{n}\vert n-1\rangle \langle n\vert\otimes \vert
-\rangle \langle -\vert,
\end{equation}
and $I_{n}$ is the identity operator in $\mathcal{H}_{n}$.

Any pure initial state can be expressed as
\begin{equation}
\vert\psi(0)\rangle=\sum_{n}\vert n\rangle \otimes [ d_{n}(0)\vert
+\rangle+e_{n}(0)\vert -\rangle],
\end{equation}
where $\lbrace \vert n\rangle \rbrace$, $n\in Z $ is an orthonormal basis
of $\mathcal{H}_{n}$ associated to the classical positions of the walker
(integer points on the line),
$\lbrace \vert +\rangle, \vert -\rangle \rbrace$ are the chirality
eigenstates in $\mathcal{H}_{S}$, and $d_{n}(0)$, $e_{n}(0)$ satisfy the
normalization condition
$\sum_{n}\left[\vert d_{n}(0)\vert^{2}+\vert e_{n}(0)\vert^{2}\right]=1$.
After $t$ applications of $U$, the QW state is
\begin{equation}\label{eqn:estadoinicialQW}
\vert\psi(t)\rangle=U^{t}\vert\psi (0)\rangle=\sum_{n}\vert n\rangle
\otimes [ d_{n}(t)\vert +\rangle+e_{n}(t)\vert -\rangle].
\end{equation}
For an infinite line, it is a well known fact that the coin system
reaches an equilibrium state as $t\rightarrow\infty$ \cite{Nayak}.
The reduced state of the coin adopts the characteristic form of
Eq.~(\ref{eqn:ARDM}), where the expression for $b$ in terms of the
wave function coefficients is
\begin{equation}
b \equiv
\begin{array}{c}
\lim_{t\rightarrow\infty}\sum_{n=-\infty }^{\infty }d_{n}(t)
e_{n}^{\ast }(t) \\
\end{array}%
.\,  \label{asym}
\end{equation}%

In Ref. \cite{Romanelli1} it is shown that the entries of the coin
RDO defined  in Eq.~(\ref{eqn:ARDM}), for an arbitrary initial
state, satisfy the relation
\begin{equation}\label{aRoma}
a(\theta,\gamma,\varphi)=\frac{{\rm {Re}}(b(\theta,\gamma,\varphi))}{\tan \theta}.\,
\end{equation}
On the other hand, Proposition (\ref{propos1}) shows that an
initial-state-dependent thermal state must verify that
\begin{equation}\label{eqn:kapppa}
\frac{b(\theta,\gamma,\varphi)}{a(\theta,\gamma,\varphi)}=\kappa(\theta),
\end{equation}
since $\kappa$ must be independent of the initial state.

Given that $b={\rm {Re}}(b)+i{\rm {Im}}(b)$ and using Eqs.~(\ref{aRoma}) and ~(\ref{eqn:kapppa}), we find
that the thermal state is guaranteed if
\begin{equation}\label{kRoma}
\left[1+\frac{i{\rm {Im}}(b(\theta,\gamma,\varphi))}{{\rm {Re}}(b(\theta,\gamma,\varphi))}\right]\tan \theta
=\kappa (\theta) ,
\end{equation}
condition that is satisfied if the real and imaginary parts of $b$ are proportional
to each other, with a proportionality constant function of only $\theta$.

As a first example, we assume that the initial state of the walker is
sharply localized at the origin, with arbitrary chirality.
Thus
\begin{equation}\label{psi0}
\vert\psi(0)\rangle=\vert 0\rangle \otimes \vert \phi_0\rangle ,
\end{equation}
where $\phi_0$ was defined in Eq.(\ref{eqn:phi0}).
This situation was studied in Ref.\cite{Romanelli1}, where an
explicit expression for $b$ was obtained (Eq.~(16) of that
reference), for a coin toss bias parameter $\theta =\pi /4$. In our
present nomenclature
\begin{equation}
b=\frac{1}{2}(1-\frac{1}{\sqrt{2}})\left[ \cos \gamma \text{ }+\sin
\gamma \text{ }(\cos \varphi +i\sqrt{2}\sin \varphi )\right] .
\label{q0entengl}
\end{equation}
In this case, from the considerations below Eq.(\ref{kRoma}), it is clear that
$\kappa$ depends on the initial state, so the equilibrium state is not an
initial-state-dependent thermal state.

As a second case, we consider strongly non-localized initial states
of the walker. One way of implementing this initial condition was
studied in Ref.\cite{Orthey}, where the following initial state is
considered
\begin{equation}
d_{n}(0)=\frac{e^{\frac{-n^{2}}{4\xi^{2}}}}{\sqrt[4]{2\pi\xi^{2}}}
\cos(\gamma/2), \label{d0}
\end{equation}
\begin{equation}
e_{n}(0)=
\frac{e^{\frac{-n^{2}}{4\xi^{2}}}}{\sqrt[4]{2\pi\xi^{2}}}\sin(\gamma/2)e^{i\varphi},
 \label{e0}
\end{equation}
which corresponds to an initially Gaussian distributed walker with
the restriction $\xi\gg1$.
We obtained the relation between $b$ and the initial condition using
directly Eqs.(19) and (29) of Ref.\cite{Orthey} in the case $\theta=\pi /4$,
\begin{equation}
b=\frac{1}{4}(\cos \gamma \text{ }+\sin \gamma \cos \varphi) .
\label{q0entengl}
\end{equation}
Following the same procedure, we have generalized this last equation
for all $\theta $. The result is
\begin{equation} \label{andr}
b=\frac{1}{2}\sin\theta\cos\alpha,
\end{equation}
where $\cos\alpha=\sin\theta\sin\gamma\cos\varphi +
\cos\theta\cos\gamma$ is the cosine of the angle determined by the
initial Bloch vector and the vector $(\sin\theta,0,\cos\theta)$.
Then, the RDO of the qubit adopts the form
\begin{small}
\begin{equation}
\overline{\rho_{_{S}}}= \frac{1}{2}\begin{pmatrix}
{1+\cos\theta\cos\alpha}&
{\sin\theta\cos\alpha}\\
{\sin\theta\cos\alpha}&
{1-\cos\theta\cos\alpha}
\end{pmatrix} .
\end{equation}
\end{small}
Since $b$ is a real number, from Eq.~(\ref{kRoma}) we conclude that
$\kappa$ is independent of the initial condition, which shows that,
in this case,  an initial-state-dependent thermal state is reached.
Additionally, the parameter $\kappa$ that defines the entanglement
Hamiltonian is, in this case,
\begin{equation}
\kappa=\tan\theta.
\end{equation}

In order to verify the geometric results of the previous section, we
calculate the eigenvalues of the coin RDO
\begin{equation}
\lambda^{\pm}=\frac{1}{2}\left(1\pm\cos\alpha\right).
\end{equation}
Therefore, according to Eq.(\ref{Tlambdas}), the entanglement temperature
is proportional to $-1/ \log\left[\tan(\alpha/2)\right]$.
The corresponding level plots over the Bloch sphere are defined by the
$\alpha$ values, or equivalently, in Cartesian coordinates, by the equation
 \begin{equation}\label{eqn:newplanes}
 x\sin\theta +z\cos\theta =\cos \alpha.
 \end{equation}
 \begin{figure}[t]
    \centering
    \includegraphics [trim= 320 80 -50 0, scale=0.5, clip]{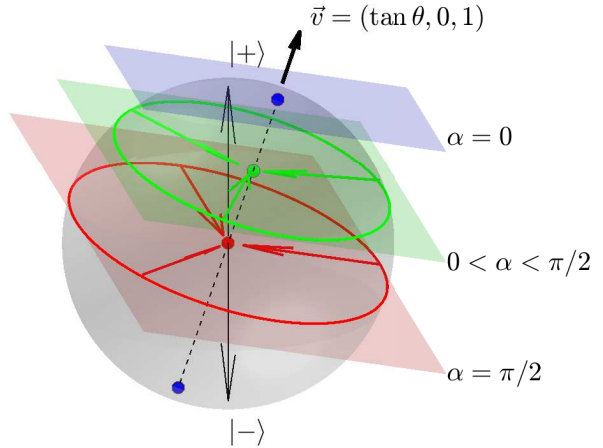}
    \caption{(Color online). Level plots of the entanglement temperature
        on the Bloch sphere, for an initially distributed Gaussian walker,
        in the limit $\xi\gg 1$.
        The initial coin states on the equator (red circumference,
        $\alpha=\pi/2$) thermalize at $T=\infty$ (red point at the center
        of the Bloch sphere), those in the intermediate latitude green
        circumference ($0<\alpha<\pi/2$)) converge to the green state at
        its center at some finite temperature, while the blue states at
        the poles, corresponding to $\alpha=1$ remainat $T=0$ during the evolution.}
    \label{fig2}
 \end{figure}
Equation (\ref{eqn:newplanes}) is a particular case of the general situation described in
Eq.(\ref{eqn:planes}), since it represents a family of planes, orthogonal to the direction $({\rm
{Re}}(\kappa),-{\rm {Im}}(\kappa),1)$ (see Fig.\ref{fig2}), determined by the parameter $\kappa=\tan\theta$.
This implies that initial qubit states located on the intersection of a particular plane with the Bloch
sphere evolve to the same asymptotic mixed state. In particular, the extreme cases $\abso{\cos\alpha}=1$
(corresponding to the initial states associated to the points $(\gamma=\theta, \varphi=0)$ and
$(\gamma=\pi-\theta,\varphi=\pi)$ (the poles defined by the privileged direction) thermalize at $T_{ent}=0$.
It is easy to see that these particular states are eigenstates of the coin operator $U_{\theta}$, and,
accordingly, entanglement never occurs, so they remain pure during the evolution, and the global state is
separable for all times $t$. This emphasizes the interpretation of the entanglement temperature as a measure
of the entanglement produced between the system and its environment.

\begin{figure}[t]
\begin{center}
\includegraphics[scale=0.6, angle=0]{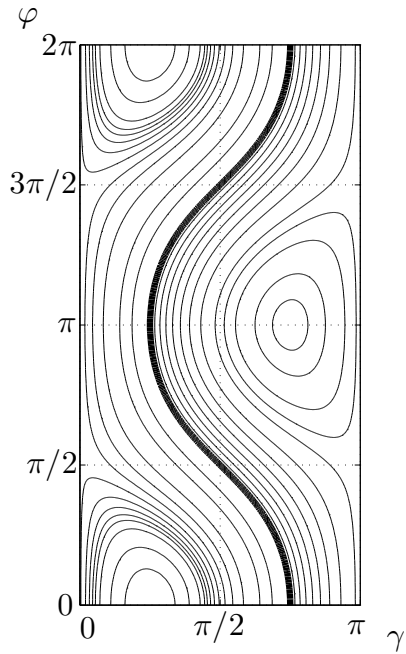}
\end{center}
\caption{Level plots of the entanglement temperature in the
($\gamma,\varphi$) space, for a Hadamard walk with an initially Gaussian
distributed walker.
The thick central line corresponds to $T=\infty$, while the center of the
closed lines are associated to the states that thermalize at $T=0$
$(\gamma=\pi /4, \varphi=0)$ and $(\gamma=3\pi /4,\varphi=\pi)$.} \label{fig3}
\end{figure}

A further verification of the previous section results is possible
analyzing the position of the thermal states on the Bloch sphere.
A particular Bloch vector $\vec{B}$ representing the coin reduced state is
\begin{equation}\label{eqn:BlochQW}
\vec{B}=\cos\alpha_{_{0}}(\sin\theta,0,\cos\theta),
\end{equation}
which satisfies $\vert\vec{B}\vert=\vert\cos\alpha_{_{0}}\vert$.
A direct calculation shows that the distance from the origin to the
corresponding level plot is $d(\vec{O},\pi_{\alpha_{0}})=\abso{\cos\alpha_{_{0}}}$.
Thus every thermal state is located at the center of its corresponding
level plot, which implies that the set of accessible thermal states is a
diameter of the Bloch sphere, as expected.
These results are illustrated in Fig.~\ref{fig2} and Fig.~\ref{fig3}.

\section{Remarks and conclusions}
Independence on the initial state is considered to be an important
requirement for thermalization in quantum systems \cite{gogolin1}.
In this work, we have shown that even in the presence of memory,
there is a class of initial-state-dependent equilibrium states for
which this dependence appears as a scalar factor in the exponent of
a thermal state expression, allowing to identify a fixed hermitian
operator that plays the role of an effective Hamiltonian, valid once
the equilibrium has reached.
These states are half way between being simple equilibrium states
and being pure thermal states, where the dependence on the initial
state disappears completely.

In the particular case of the thermal state, the effective Hamiltonian
operates as a magnetic field- spin interaction.
It is determined by a dimensionless parameter $\kappa$ that depends
on the relevant physical parameters involved in  the global dynamics,
selecting a privileged direction along which the thermal states are located.
In the particular case of the QW on the line, $\kappa$ is the tangent of
the coin bias, $\kappa=\tan\theta$.

The initial-state-dependent entanglement temperature associated to the
defined thermal states can serve as a measure of the entanglement produced
between the subsystem and its environment, as illustrated in the case of a QW.

It must be emphasized the role of a large initial bath occupation for this
kind of thermalization to occur.
In the case of  a localized walker, the system evolves to an equilibrium
state that cannot be written in the exponential form with a fixed Hamiltonian,
{\it i.e.}, one valid for all initial states of the coin.
This seems to agree with previous results, {\it e.g.} of Ref.\cite{Linden}.

It is remarkable that although the example employed in this work to
illustrate the theoretical results does not correspond to a typical
thermal contact process, but a rather abstract one, the thermalization
obtained is identical to that found in other physical systems \cite{Romanelli2}.
This suggests that a more profound analysis of the underlying mathematical
structure could reveal some kind of universality, aspect that needs to be explored.
The study of other two-level open systems from the perspective presented
in this work could be interesting, and  it is currently under investigation.
\section*{Acknowledgments}
 This work was partially supported by ANII and PEDECIBA (Uruguay).


\begin{thebibliography}{99}
    \bibitem{Schulman}L. S. Schulman. Phys. Rev. A \textbf{18}, 2379 (1978).
\bibitem{Gemmer}J. Gemmer, M. Michel, and G. Mahler, Quantum Thermodynamics:
The Emergence of Thermodynamical Behavior within Composite Quantum Systems.
Lecture Notes in Physics, Springer (2004).
\bibitem{Parrondo} Juan M.R. Parrondo, Jordan M. Horowitz and Takahiro Sagawa,
Nature Physics \textbf{11}, 131 (2015).
\bibitem{Goold} John Goold, Marcus Huber, Arnau Riera, Lidia del Rio and Paul Skrzypczyk,
J. Phys. A: Math. Theor \textbf{49}, 143001 (2016).
\bibitem{Polkovnicov} Anatoli Polkivnikov and Dries Sels,  Science \textbf{353}, 752 (2016).
\bibitem{Kaufman} Adam M. Kaufman, M. Eric Tai, Alexander Lukin, Matthew Rispoli,
Robert Schittko, Philipp M. Preiss and Markus Greiner,  Science \textbf{353}, 794 (2016).
\bibitem{Romanelli1} A. Romanelli, Phys. Rev. A \textbf{85}, 012319 (2012).
\bibitem{Romanelli2} A. Romanelli, R. Donangelo, A. Vallejo, Physica A  \textbf{437}, 471(2015).
\bibitem{Novotny} M. A. Novotny, F. Jin, S. Yuan, S. Miyashita, H. De Raedt, and K. Michielsen,
Phys. Rev. A \textbf{93}, 032110 (2016).
\bibitem{schliemann} J. Schliemann, J. Stat. Mech. \textbf{09}, P09011 (2014).
\bibitem{Linden} Linden, N., S. Popescu, A. Short, and A. Winter, Phys.
Rev. E \textbf{79}, 061103 (2009)
\bibitem{Goldstein} S. Goldstein, J. Lebowitz, R. Tumulka, and N. Zanghi,
Phys.Rev.Lett.  \textbf{96}, 050403 (2006).
\bibitem{Popescu} S. Popescu, A. Short, A. Winter,  Nature Physics \textbf{2}, 754 (2006).
\bibitem{Nielsen} M.A. Nielsen, I.L. Chuang, Quantum Computation and Quantum Information,
Cambridge University Press, Cambridge (2000).
\bibitem{Bocchieri} P.Bocchieri, A. Loinger, Phys. Rev. \textbf{107} (1957).
\bibitem{kossloff}R. Kosloff, Entropy \textbf{15}, 2100 (2013).
\bibitem{travaglione}  B. Travaglione,  G. Milburn,  Phys. Rev. A \textbf{65}, 032310 (2002).
\bibitem{Renato} Renato Portugal, Quantum Walks and Search Algorithms. Springer, New York, (2013).
\bibitem{Grover} L. K. Grover. Proceedings of the 28th ACM Symposium on the
Theory of Computing, 212 (1996), quant-ph/9605043 (1996).
\bibitem{Nayak}A. Nayak, A. Vishvanath,  quant-ph/0010117, (2000).
\bibitem{Carneiro} I. Carneiro, M. Loo, X. Xu, M. Girerd, V. M. Kendon, and
P. L. Knight, New J. Phys. \textbf{7}, 56 (2005).
\bibitem{Abal} G. Abal, R. Siri, A. Romanelli, and R. Donangelo, Phys. Rev.
A \textbf{73}, 042302, 069905(E) (2006).
\bibitem{Annabestani} M. Annabestani, M. R. Abolhasani and, G. Abal, J.
Phys. A: Math. Theor. \textbf{43}, 075301 (2010).
\bibitem{Omar} Y. Omar, N. Paunkovic, L. Sheridan, and S. Bose, Phys. Rev.
A, \textbf{74}, 042304 (2006)
\bibitem{Pathak} P. K. Pathak, and G. S. Agarwal, Phys. Rev. A, \textbf{75},
032351 (2007)
\bibitem{Petulante} C. Liu, and N. Petulante, Phys. Rev. A \textbf{79},
032312 (2009).
\bibitem{Venegas} S. E. Venegas-Andraca, J.L. Ball, K. Burnett, and S. Bose,
New J. Phys., \textbf{7}, 221 (2005).
\bibitem{Endrejat} J. Endrejat, H. B\"{u}ttner, J. Phys. A: Math. Gen.
\textbf{38}, 9289 (2005).

\bibitem{Ellinas1} A.J. Bracken, D. Ellinas, and I. Tsohantjis, J. Phys. A:
Math. Gen. \textbf{37}, L91 (2004).

\bibitem{Ellinas2} D. Ellinas, and A.J. Bracken, Phys. Rev. A \textbf{78},
052106 (2008).

\bibitem{Maloyer} O. Maloyer, and V. Kendon, New J. Phys., \textbf{9}, 87
(2007).

\bibitem{gogolin1} C. Gogolin and J. Eisert,
Reports on Progress in Physics \textbf{79},  056001 (2016).
\bibitem{Orthey} A. C. Orthey, E. P. M. Amorim, Quantum Inf. Process. \textbf{16}, 224 (2017).
\end{thebibliography}
\end{document}